\documentclass{ptephy_v1d}
\preprintnumber{KEK-TH-1572, J-PARC-TH-0015} 
\usepackage{graphicx,amsmath}
\newcommand{\ra}{\rightarrow}
\def\lsim{\mathrel{\rlap{\lower4pt\hbox{\hskip1pt$\sim$}}
    \raise1pt\hbox{$<$}}}

\usepackage{color}

\definecolor{mycolor}{rgb}{0.6,0.0,0.4}

\begin{document}

\title{Impacts of B-factory measurements on 
determination of fragmentation functions \\
from electron-positron annihilation data}

\author{
\name{M. Hirai}{1,2}, 
\name{H. Kawamura}{3,4}, 
\name{S. Kumano}{4,5}, and
\name{K. Saito}{2}
}

\address{
\affil{1}{$^1$Nippon Institute of Technology, Saitama 345-8501, Japan} \\
\affil{2}{$^2$Department of Physics, Faculty of Science and Technology,
             Tokyo University of Science, 
             2641, Yamazaki, Noda, Chiba, 278-8510, Japan} \\
\affil{3}{$^3$Department of Mathematics, Juntendo University, Inzai,
          Chiba 270-1695, Japan} \\
\affil{4}{$^4$KEK Theory Center,
             Institute of Particle and Nuclear Studies,
             High Energy Accelerator Research Organization (KEK),
             1-1, Ooho, Tsukuba, Ibaraki, 305-0801, Japan} \\           
\affil{5}{$^5$J-PARC Branch, KEK Theory Center,
             Institute of Particle and Nuclear Studies, KEK 
           and
           Theory Group, Particle and Nuclear Physics Division, 
           J-PARC Center,
           203-1, Shirakata, Tokai, Ibaraki, 319-1106, Japan} \\
}


\begin{abstract}
Fragmentation functions are determined for the pion and kaon
by global analyses of charged-hadron production data in
electron-positron annihilation. Accurate measurements were 
reported by the Belle and BaBar collaborations 
for the fragmentation functions at the center-of-mass energies of 
10.52 GeV and 10.54 GeV, respectively, at the KEK and SLAC B factories,
whereas other available $e^+e^-$ 
measurements were mostly done at higher energies, mainly at the $Z$ mass
of 91.2 GeV. There is a possibility that gluon fragmentation functions, 
as well as quark fragmentation functions, are accurately determined 
by scaling violation. We report our global analysis of the fragmentation
functions especially to show impacts of the B-factory measurements 
on the fragmentation function determination. Our results indicate 
that the fragmentation functions are determined more accurately
not only by the scaling violation but also by high-statistical 
nature of the Belle and BaBar data.
However, there are some tensions between the Belle and BaBar data
in comparison with previous measurements.
We also explain how the flavor dependence of
quark fragmentation functions and the gluon function are separated
by using measurements at different $Q^2$ values.
In particular, the electric and weak charges are different depending
on the quark type, so that a light-quark flavor separation also 
became possible in principle due to the precise data 
at both $\sqrt{s}\simeq 10.5$ GeV and 91.2 GeV.
\end{abstract}

\subjectindex{B65, C22, D32}
\maketitle

\section{Introduction}\label{intro}

Semi-inclusive hadron production processes become increasingly important
for probing hadrons and quark-gluon system properties and
for finding a signature beyond the standard model
in high-energy hadron reactions. 
For describing their cross sections, there are
three ingredients: parton distribution functions (PDFs) of 
initial hadrons, intermediate partonic interactions, and 
fragmentation functions in the final state.
The PDFs have been well investigated in a wide kinematical
region with a variety of experimental measurements,
and the partonic interactions can be calculated in perturbative
QCD (Quantum Chromodynamics). However, the fragmentation functions
have not been accurately determined as it was typically shown in 
Ref. \cite{hkns07}. Namely, there are large uncertainty bands
especially in disfavored-quark and gluon functions even in the pion,
for which relatively accurate data exist.
Moreover, the error bands are large for all the functions in the kaon 
and nucleon. This fact should be kept in mind for drawing any physics 
conclusions from experimental measurements of high-energy 
hadron-production processes \cite{high-pt}.

The fragmentation functions are obtained by a global analysis
of experimental data on the hadron productions 
\cite{hkns07,ffs_before_2006,ffs-summary,ffs-recent,Epele:2012vg}
in the similar way to the determination of the PDFs.
Although some low moments of the PDFs can be calculated
in lattice QCD, it is not possible to calculate the fragmentation
functions and their moments 
due to a specific final state.
There are some hadron models to calculate the fragmentation functions
\cite{ff-model}; however, they are
not accurate enough to calculate precise cross sections for
various processes. Therefore, the global analysis of
world experimental data is the most reliable way to
obtain the accurate fragmentation functions.
Furthermore, nuclear modifications of the fragmentation functions 
are also discussed recently \cite{nuclear-ffs}
for understanding heavy-ion reactions and semi-inclusive lepton 
deep inelastic scattering from nuclei \cite{hermes-nuclear}.

The fragmentation functions have been determined by several groups
from analyses of experimental data on hadron productions. Until 2013,
the center-of-mass (c.m.) energy of the hadron production process 
$e^+ +e^- \rightarrow h+X$ ranged mostly from 12 GeV to 91.2 GeV;
however, many data were taken 
in the $Z$ mass region of 91.2 GeV at SLD (SLAC Large Detector) and 
CERN-LEP (Large Electron-Positron Collider).
There were lower-energy data, for example at 
12, 14, 22, 29, 30, 34, 44, and 58 GeV;
however, they are taken in a limited kinematical region
as shown in Fig. 1 of Ref. \cite{hkns07}.
This fact suggests that significant scaling violation should not be
found in the data, which leads to large error bands in
gluon fragmentation functions \cite{hkns07},
whereas the gluon distribution function has been determined 
in the nucleon by scaling violation data mainly taken at
HERA (Hadron-Electron Ring Accelerator).

In 2013, there were new experimental developments on the fragmentation 
functions in the sense that very accurate data were obtained by 
the Belle and BaBar collaborations for the pion and kaon \cite{belle,babar}
at the energies of 10.52 GeV and 10.54 GeV, respectively.
They are measured in the wide momentum-fraction region 
which was not covered by the previous measurements. 
These $e^+ +e^- \rightarrow h+X$ experiments
were done at the B factories of KEK and SLAC.
It is particularly important that the measurements are
at a much lower energy than the $Z$ mass
because the scaling violation becomes clear
in the fragmentation functions.
Then, the gluon fragmentation functions should be determined
more accurately. It is known that the gluon functions are
very important in analyzing hadron production processes
especially at RHIC (Relativistic Heavy Ion Collider) and 
LHC (Large Hadron Collider),
where gluons play a major role in the productions.
In addition, the high-statistical B-factory data should be
also valuable to obtain precise quark and antiquark 
fragmentation functions.

These considerations made us to investigate the role of the B-factory
data in the determination of the fragmentation functions because our
functions are determined in 2007 without these data \cite{hkns07}.
A purpose of this work is to show how error bands of 
the fragmentation functions are reduced by adding the B-factory data,
as shown in Sec. \ref{results}.
In addition, we discuss the details of possible flavor separation
and determination of gluon function by taking advantage of 
accurate measurements in the wide kinematical range between
10.5 GeV to 91.2 GeV, as explained in Sec. \ref{flavor}.
The scaling violation should play an important
role in determining the gluon fragmentation function.

This paper is organized as follows. 
In Sec. \ref{formalism}, hadron-production cross sections
and the fragmentation functions are introduced 
in the $e^+e^-$ annihilation.
Then, our global analysis method is explained in Sec. \ref{analysis}
for determining the optimum functions, and results are
discussed in Sec. \ref{results}. Finally, this work is summarized
in Sec. \ref{summary}.

\vspace{-0.0cm}
\section{Formalism}
\label{formalism}

The total fragmentation function is defined 
by the hadron-production cross section for 
electron-positron annihilation ($e^+ + e^- \rightarrow h + X$) 
and the total hadronic cross section $\sigma_{tot}$ \cite{esw-book}:
\begin{equation}  
F^h(z,Q^2) = \frac{1}{\sigma_{tot}} 
\frac{d\sigma (e^+e^- \rightarrow hX)}{dz} .
\label{eqn:def-ff}
\end{equation}
Here, the $Q^2$ is given by the c.m. energy $\sqrt{s}$ as
$Q^2=s$, and 
the variable $z$ is defined by the energy fraction: 
\begin{equation}   
z \equiv \frac{E_h}{\sqrt{s}/2} = \frac{2E_h}{\sqrt{Q^2}},
\label{eqn:def-z}
\end{equation}
where $E_h$ and $\sqrt{s}/2$  are the hadron and beam energies,
respectively. 
The process $e^+ +e^- \rightarrow h+X$ is described by two steps. 
First, a quark-antiquark pair is created by 
$e^+ e^- \rightarrow q\bar q$ as shown in Fig. \ref{eqn:def-ff},
where the intermediate state is virtual $\gamma$ or $Z$.
Therefore, the variable $Q^2$ is the virtual photon or $Z$ momentum squared
in $e^+e^- \rightarrow \gamma, Z$.
Higher-order corrections such as $e^+ e^- \rightarrow q\bar q g$
are taken into account in the NLO analysis.
Second, a hadron $h$ is created from quark ($q$), antiquark ($\bar q$),
or gluon ($g$), and this process is called fragmentation.
The fragmentation function approximately indicates  
the probability for producing a hadron, 
and its universality is essential for describing any hadron-production
processes at high-energy reactions.
The total cross section $\sigma_{tot}$ in Eq. (\ref{eqn:def-ff})
is described by the $q\bar q$-pair creation processes,
$e^+e^- \rightarrow \gamma \rightarrow q\bar q$ and
$e^+e^- \rightarrow Z \rightarrow q\bar q$, with additional
higher-order corrections. The explicit expression of $\sigma_{tot}$ 
is found in Ref. \cite{qqbar-cross}.

\begin{figure}[b]
\vspace{0.0cm}
\begin{center}
   \includegraphics[width=5.0cm]{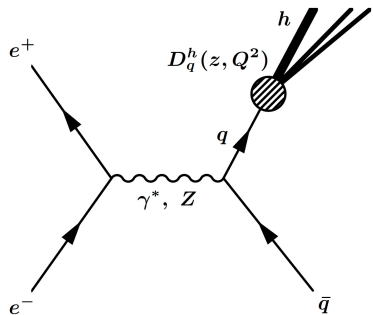}
\end{center}
\vspace{-0.5cm}
\caption{Hadron production in electron-positron annihilation
         ($e^+ + e^- \rightarrow h + X$).}
\label{fig:ee-ff}
\end{figure}

A fragmentation process from a quark $q$ is shown in Fig. \ref{fig:ee-ff}
as an example; however, the fragmentation occurs from any primary quarks, 
antiquarks, and gluons. Therefore, the function $F^h(z,Q^2)$ is expressed
by the sum of their contributions:
\begin{align}  
F^h(z,Q^2) & = \sum_i \int^{1}_{z} \frac{dy}{y}
                      C_i(z/y,\alpha_s) \, D_i^h (y,Q^2)
            \equiv \sum_i C_i(z,\alpha_s) \otimes D_i^h (z,Q^2) ,
\label{eqn:def-ffqqbarg}
\end{align}
where the fragmentation function of the hadron $h$ from a parton $i$
($=u,\ \bar{u}, \ d,\ \bar{d},\ \cdot\cdot\cdot,\ g$) is denoted as 
$D_i^h(z,Q^2)$,
and a coefficient function is $C_i(z,\alpha_s)$ which contains
perturbative QCD corrections and electroweak coupling constants.
The notation $\otimes$ indicates a convolution integral defined by
$ f (x) \otimes g (x) = \int^{1}_{x} dy/y  f(y) g(x/y) $.
Formally, the fragmentation function for the $i$-th quark 
is expressed as \cite{ffs-def}
\begin{align}
\! 
D_i^h (z) = & \sum\limits_X \int \frac{dy^-}{12 \pi} e^{ik^+ y^-}
{\rm{Tr}} \left[ {\gamma^+  
\left\langle {0 \left| {\left. {\left. \psi _i (0,y^-  ,0_\bot) \right|h,X} 
\right\rangle  } \right.} \right.} \right.
\! \! \! \! 
\left.  {\left. {\left. { 
\left\langle {h,X\left| {\bar \psi _i (0)} \right.} \right.} 
\right|0} 
\right\rangle } \right] ,
\label{eqn:formal-ffs}
\end{align}
where $k$ is the parent quark momentum, and
the variable $z$ is given by the momentum ratio
$z=p_h^+/k^+$ with the hadron momentum $p_h$.
Lightcone variables are defined by $a^\pm =(a^0 \pm a^3)/\sqrt{2}$, 
and $\perp$ is the direction perpendicular to the third coordinate.
To be precise, a gauge link is needed in Eq. (\ref{eqn:formal-ffs})
for the color gauge invariance. 
The equation indicates that a specific hadron $h$ should 
be observed in the final state with the momentum fraction $z$.
It suggests that the operator-product-expansion method 
cannot be applied, which is the reason why 
the fragmentation functions are not evaluated in lattice QCD.

The scaling violation, namely the $Q^2$ dependence, of 
the fragmentation functions (FFs) is described 
by the DGLAP (Dokshitzer-Gribov-Lipatov-Altarelli-Parisi) evolution equation. 
Its general form is the integro-differential equation \cite{esw-book}
\begin{align}
& \! \! \! \! 
\frac{\partial}{\partial \ln Q^2}
\left(
\begin{array}{c}
D_{f}^h (z,Q^2) \\
D_{g}^h (z,Q^2)
\end{array}
\right) \! = \frac{\alpha_s (Q^2)}{2\pi} \! \! 
  \sum_{f'=q_j,\bar{q}_j} \! \!
   \left(
\begin{array}{cc}
P_{f' f}(z,\alpha_s) &   P_{g f}(z,\alpha_s) \\
P_{f' g}(z,\alpha_s)     &   P_{g g}(z,\alpha_s) 
\end{array}
   \right)
\otimes 
\left(
\begin{array}{c}
D_{f'}^h (z,Q^2) \\
D_{g}^h (z,Q^2) 
\end{array}
\right) ,
\label{eqn:evolution}
\end{align}
where $f$ indicates $f=q_i, \bar q_i$,
$N_f$ is the number of quark flavors,
$\alpha_s (Q^2)$ is the running coupling constant of QCD,
and the function $P_{ij}(z)$ is time-like splitting function.
One notices that $i$ and $j$ are interchanged from the space-like
DGLAP equations for the PDFs. In addition,
it should be noted that the time-like splitting functions are
generally different from the space-like ones if higher-order
perturbative corrections are taken into account
\cite{esw-book,splitting,space-time}. The actual expressions 
of $P_{ij}(z)$ are lengthy in the NLO (next-to-leading-order) 
of $\alpha_s$, so that they should be found, for example, 
in Ref. \cite{esw-book}.
  
The evolution equations are complicated integro-differential
equations especially if higher-order $\alpha_s$ corrections
are included. Therefore, they cannot be solved in analytical methods.
Various numerical methods have been developed and their references
should be found in Refs. \cite{evolution,hk-q2evol}.
In the work of Ref. \cite{hk-q2evol}, the equations are
solved by using the Gauss-Legendre quadrature for evaluating integrals,
and a useful code is provided for calculating the $Q^2$ evolution 
of the fragmentation functions in the leading order (LO) 
and NLO of $\alpha_s$. The renormalization scheme is 
the modified minimal subtraction ($\overline {\rm MS}$) scheme
in the NLO evolution. We use this $Q^2$ evolution code 
in our global analysis of this article.

\section{Analysis method}
\label{analysis} 

\subsection{Initial fragmentation functions}
\label{initial}

The FFs are determined by global analyses of world data on
hadron-production processes. They are parametrized in a simple
polynomial form of $z$ at a fixed $Q^2$ which is denoted as $Q_0^2$.
Because it is the purpose of this work to show modifications
of uncertainties in the FFs from the previous analysis 
without the B-factory data \cite{hkns07}, we use the same function
\begin{equation}
D_i^h(z,Q_0^2) = N_i^h z^{\alpha_i^h} (1-z)^{\beta_i^h} ,
\end{equation}
where $N_i^h$, $\alpha_i^h$, and $\beta_i^h$ are parameters
to be determined by the global analysis.
A more complicated functional form is usually employed in the PDF analysis;
however, the data variety is enough to probe minute $z$ dependence
in the FF case.
In our analysis, the initial scale $Q_0^2$ is taken as $Q_0^2=1$ GeV$^2$ 
for gluon and light quarks ($u$, $d$, $s$), and it is
taken at the masses $m_c^2$ and $m_b^2$ for charm
and bottom quark FFs. 

The second moments of the FFs are defined by
\begin{equation}
M_i^h = \int_0^1 dz \, z \, D_i^h (z,Q^2) ,
\label{eqn:2n}
\end{equation}
and they are related to the overall constants $N_i^h$ 
with the beta function as
$N_i^h = M_i^h / B(\alpha_i^h+2, \beta_i^h+1)$.
Since there is a sum rule for the second moments: $\sum_h M_i^h =1$,
it is more appropriate to use the parameter set
($M_i^h$, $\alpha_i^h$, $\beta_i^h$)
in the global analysis, rather than the set
($N_i^h $, $\alpha_i^h$, $\beta_i^h$)
in order to exclude an unphysical solution 
with a sum which significantly exceeds one 
even by the summation over $h=\pi$ and $K$.

In general, different parameters are assigned for favored and
disfavored FFs, separately.
The favored means the fragmentation from a quark which exits 
in the hadron $h$ as a constituent in the naive SU(6) quark model,
whereas the disfavored means the fragmentation from a sea quark.
Although it is known that light sea-quark 
(up, down, strange sea quarks) distributions are not
flavor symmetric in the unpolarized PDFs \cite{flavor3},
they are assumed to be the same 
in the present analysis of the FFs.

Considering the basic quark configuration in the pion ($\pi^+ (u \bar d)$),
we assign the same functions for the favored functions
of $u$ and $\bar d$:
\begin{equation}
D_{u}^{\pi^+} (z,Q_0^2)  = D_{\bar d}^{\pi^+} (z,Q_0^2) .
\label{favored}
\end{equation}
On the other hand, the disfavored functions of $\pi^+$ are
assumed to be equal in the initial scale:
\begin{align}
D_{\bar u}^{\pi^+} (z,Q_0^2) & = D_{d}^{\pi^+} (z,Q_0^2)
  = D_{s}^{\pi^+} (z,Q_0^2)
  = D_{\bar s}^{\pi^+} (z,Q_0^2) .
\label{disfavored}
\end{align}
In addition, we have separate gluon, charm-quark, and bottom-quark FFs:
\begin{align}
D_{g}^{h} (z,Q_0^2) , \ \ 
D_{c}^{h} (z,m_c^2) = D_{\bar c}^{h} (z,m_c^2) , \ \ 
D_{b}^{h} (z,m_b^2) = D_{\bar b}^{h} (z,m_b^2) ,
\label{eqn:gcb}
\end{align}
where $h=\pi^+$ or $K^+$ in the following kaon parametrization.
In analyzing the pion data of the $e^+e^-$ annihilation,
the charged-pion combination ($\pi^+ +\pi^-$) data are analyzed.
The FFs of $\pi^-$ are obtained from the $\pi^+$ ones by
using the charge symmetry at any $Q^2$ which is not
necessarily $Q_0^2$ as long as $Q^2$ is in the perturbative QCD region:
\begin{align}
D_{q}^{\bar h} (z,Q^2) 
= D_{\bar q}^{h} (z,Q^2) , 
\ \ 
D_{g}^{\bar h} (z,Q^2)
= D_{g}^{h} (z,Q^2)  .
\label{eqn:hbar}
\end{align}

The FFs of the kaon are taken in the same way by considering
the constituent-quark configuration $K^+ (u \bar s)$; however,
the antistrange function is taken differently from the up-quark
function in the favored FFs:
\begin{align}
D_{u}^{K^+} (z,Q_0^2), \ \ D_{\bar s}^{K^+} (z,Q_0^2) .
\end{align}
The disfavored functions are taken as the same:
\begin{align}
D_{\bar u}^{K^+} (z,Q_0^2)
& = D_{d}^{K^+} (z,Q_0^2)
= D_{\bar d}^{K^+} (z,Q_0^2)
= D_{s}^{K^+} (z,Q_0^2)  .
\end{align}
There are also gluon, charm-quark, and bottom-quark FFs
as given in Eq. (\ref{eqn:gcb}).
For $K^-$, the relations of Eq. (\ref{eqn:hbar}) are used 
to obtain the functions from the ones of $K^+$.

\subsection{Experimental data}
\label{data}

\begin{table}[b]
\caption{Experimental collaborations, laboratories, references, 
     center-of-mass energies, and numbers of data points are listed 
     for used data sets of 
     $e^+ +e^- \rightarrow \pi^\pm +X$, $K^\pm +X$ \cite{durham}.
     TASSO $\sqrt{s}=44$ GeV data exist only for $\pi^\pm$.}
\label{tab:exp-pion-kaon}
\centering
\begin{tabular}
{l@{\extracolsep{10ptplus1fil}}c@{\extracolsep{10ptplus1fil}}c
@{\extracolsep{10ptplus1fil}}c@{\extracolsep{10ptplus1fil}}c
@{\extracolsep{10ptplus1fil}}c}
Experiment  &  Lab. \ & Ref.  & $\sqrt{s}$ \ & \# of $\pi$ data & \# of K data \\
\hline
Belle            &  KEK \ & \cite{belle}    & 10.52          &  78  &  78  \\
BaBar            &  SLAC \ & \cite{babar}   & 10.54          &  36  &  36  \\
TASSO            &  DESY  & \cite{tasso12_30,tasso14_22,tasso34_44}
                                       & 12,14,22,30,34,(44) &  29  &  18  \\
TPC              &  SLAC  & \cite{tpc29}    & 29             &  18  &  17  \\
HRS              &  SLAC  & \cite{hrs29}    & 29             & \, 2 & \, 3 \\
TOPAZ            &  KEK \ & \cite{topaz58}  & 58             & \, 4 & \, 3 \\
SLD              &  SLAC  & \cite{sld91}    & 91.28          &  29  &  29  \\
SLD ($u,d,s$)    &  SLAC  & \cite{sld91}    & 91.28          &  29  &  29  \\
SLD ($c$)        &  SLAC  & \cite{sld91}    & 91.28          &  29  &  29  \\
SLD ($b$)        &  SLAC  & \cite{sld91}    & 91.28          &  29  &  28  \\
ALEPH            &  CERN  & \cite{aleph91}  & 91.2  \,       &  22  &  18  \\
OPAL             &  CERN  & \cite{opal91}   & 91.2  \,       &  22  &  10  \\
DELPHI           &  CERN  & \cite{delphi91} & 91.2  \,       &  17  &  27  \\
DELPHI ($u,d,s$) &  CERN  & \cite{delphi91} & 91.2  \,       &  17  &  17  \\
DELPHI ($b$)     &  CERN  & \cite{delphi91} & 91.2  \,       &  17  &  17  \\
\hline
total            &        &                 &             & 378 \, & 359 \, \\
\end{tabular}
\end{table}

The fragmentation functions are determined by global analyses 
of charged-hadron production data of $e^+ +e^- \rightarrow h^\pm +X$,
where the hadron $h^\pm$ is charged pion ($\pi^+ + \pi^-$)
or charged kaon ($K^+ + K^-$) in this work.
We use the data from the measurements of
TASSO \cite{tasso12_30,tasso14_22,tasso34_44},
TPC \cite{tpc29}, HRS \cite{hrs29}, TOPAZ \cite{topaz58}, 
SLD \cite{sld91}, ALEPH \cite{aleph91}, OPAL \cite{opal91}, and 
DELPHI \cite{delphi91,delphi91-2} as they are employed 
in the analysis of 2007 \cite{hkns07}.
In addition, we include the new Belle \cite{belle} and BaBar \cite{babar}
data in our analyses of pion and kaon.
Experimental collaborations, laboratories,
references, center-of-mass (c.m.) energies, and numbers of the data are
listed in Table \ref{tab:exp-pion-kaon} \cite{durham}.

The kinematical cuts, $Q^2 >1$ GeV$^2$, 
$z>0.1$ ($z>0.15$) for the data at $\sqrt{s}<M_Z$ (BaBar kaon), and
$z>0.05$ for the data at $\sqrt{s}=M_Z$,
are applied in using the measured data. 
The perturbative QCD is used in obtaining the coefficient functions
and splitting functions in Eqs. (\ref{eqn:def-ffqqbarg}) and
(\ref{eqn:evolution}), respectively.
$Q^2 >1$ GeV$^2$ is considered to be region where the perturbative QCD
can be applied. The $z$-cut condition is applied because of 
resummation of soft-gluon logarithms \cite{resum}.
The resummation effects need to be properly handled in the formalism
for describing the small-$z$ data, whereas a fixed-order formalism
is used in this work.
We applied the $z$ cut for the BaBar data at $z > 0.15$ because 
the total fragmentation function decreases steeply at small $z$, 
where the modified leading logarithm approximation (MLLA) 
is needed for theoretically describing 
such behavior \cite{babar,resum}. In addition,
the BaBar multiplicities are shown by the momentum fraction $x_p$, and
the MLLA-necessity region could shift to larger $z$ ($\approx 0.10-0.15$) 
if the finite kaon mass is used in converting $x_p$ to $z$. 
Since we use the fixed-order formalism in analyzing the data,
these small-$z$ cut conditions are applied.

\begin{figure}[t]
\vspace{-0.0cm}
\begin{center}
\includegraphics[width=0.60\textwidth]{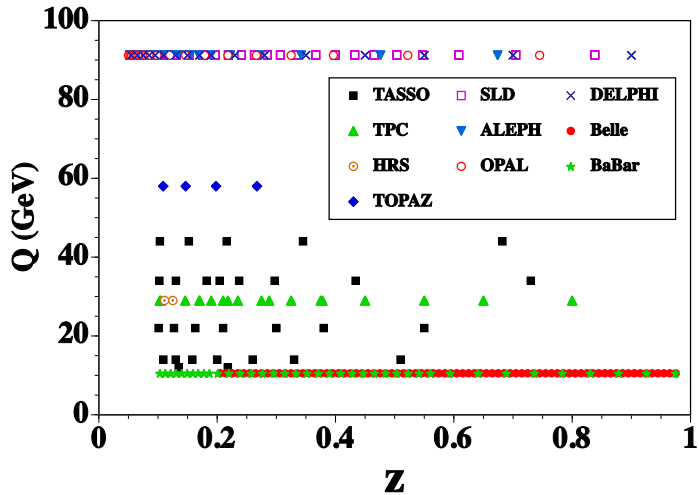}
\end{center}
\vspace{-0.5cm}
\caption{Kinematical range is shown by $z$ and $Q (=\sqrt{s})$ values 
         for the pion data. The Belle and BaBar data are shown 
         by the filled circles and asterisks at $Q \simeq 10.5$ GeV, 
         respectively.}
\label{fig:zq2}
\end{figure}

The kinematical region of the pion data is shown in Fig. \ref{fig:zq2}.
In comparison with the previous analysis \cite{hkns07},
the additional data are from the Belle and BaBar  measurements, which are
shown by the filled circles and asterisks 
at $Q=10.52$ GeV and 10.54 GeV. It is obvious that these B-factory data
significantly extend the kinematical region at small $Q^2$ at large $z$.
As one notices from the figure, many previous data are taken especially
at the $Z$ mass $M_Z$. There are not so many data in the lower-energy 
region except for the Belle and BaBar. The kaon data of the Belle and BaBar
collaborations are taken in the same kinematical region.

The B-factory data are taken at 10.5 GeV. In comparison with the data 
at $M_Z$, the scaling violation should become clearer for the first time 
in the fragmentation functions. Since the scaling violation is used 
for determining the gluon distribution in the PDFs, it became realistic 
to determine the gluon fragmentation functions by the B-factory data
in combination with others. In using the Belle data, one needs to 
be careful about the correction factor due to the initial-state radiation
(ISR). Because of the ISR, the energy scale $\sqrt{s}/2$ is modified
and the measurements contain the variation of this scale. 
The Belle data are supplied by applying a cut for the energy variation
$E_{\text{ISR}}<0.5\% \sqrt{s}/2$, namely
$ (d\sigma^h /dz)_{E_{\text{ISR}}<0.5\% \sqrt{s}/2}
  = c \cdot d\sigma^h /dz $,
where $c$ is the normalization correction factor due to the kinematical cut.
Therefore, if the total cross section is estimated by the Belle data
in a global analysis, the total fragmentation function should be given by
\begin{align}
F^h
= \frac{1}{(\sigma_{tot})_{E_{\text{ISR}}<0.5\% \frac{\sqrt{s}}{2}}}
  \left ( \frac{d\sigma_{e^+e^- \rightarrow hX}}{dz} 
  \right )_{E_{\text{ISR}}<0.5\% \frac{\sqrt{s}}{2}}
= \frac{1}{c \, \sigma_{tot}}
   \left ( \frac{d\sigma_{e^+e^- \rightarrow hX}}{dz} 
  \right )_{\text{Belle data}} ,
\end{align}
where the total cross section $\sigma_{tot}$ is calculated theoretically.
The correction factor $c$ is estimated by the Monte-Carlo simulation
by the Belle collaboration. It could be approximated by an $z$-independent
constant, and it is given by $c=0.64616 \pm 0.00003$ \cite{belle}.

The BaBar measurements are provided by two types of data:
conventional and prompt. The prompt hadron data include primary hadrons
or decay products from particles with lifetimes shorter than $10^{-11}$s.
This set does not include the decays of hadron with 
lifetimes in the range $1-3 \times 10^{-11}$s, such as
weakly decaying baryons and $K_S^0$. These decays are added
to the prompt data and the resulting data set is called
the conventional one. Most other measurements 
include all the decays, so that the conventional set is
close to the other publications obtained by similar data handling.
In our work, the conventional set is used; however, we find that
the prompt set is numerically more consistent with previous measurements
of the LEP and SLD. In comparison with the Belle pion data,
we find that the Belle measurements are between the BaBar prompt and 
conventional data sets. The Belle pion data are larger than the BaBar
prompt in the total fragmentation function of Eq. (\ref{eqn:def-ff}), 
so that a large normalization correction is needed for 
the Belle if the BaBar prompt data are included in the global analysis.
Since there is no such significant difference between the prompt
and conventional sets in the kaon, the normalization 
shifts for both Belle and BaBar data are not large as we show 
in Sec. \ref{results}.

\subsection{Flavor separation and determination of gluon fragmentation function}
\label{flavor}

The total FF measured in $e^+e^-$ annihilation 
is given in Eqs.(\ref{eqn:def-ff}) and (\ref{eqn:def-ffqqbarg}).
Since the contribution from a quark and its antiquark 
is equal, the cross section is given by  
the ``plus-component"of the quark FFs, 
$D_{q_i^+}\equiv D_{q_i}+D_{\bar{q}_i}$, and the gluon FF $D_g$.
In the global analysis, the heavy-flavor FFs are mainly determined 
from the flavor-tagged data \cite{sld91,delphi91-2}, 
while separation of the light-flavor FFs is possible due to  
the charge difference between the electromagnetic and weak interactions
and also due to the effects of the scale evolution with a help of some 
plausible assumptions for the favored and disfavored FFs 
at the initial scale.
The gluon FF is determined mainly through the scale evolution. 

To illustrate these points,   
let us consider the pure-QED process at $Q^2\ll M_z^2$,
where the intermediate $Z$ boson can be neglected.
For simplicity, we take the number of flavor
$N_f=3$ in the following discussion
or we consider the cross sections 
by subtracting the heavy-quark contributions.
In any case, the charm and bottom functions could be determined
separately by the $c$- and $b$-tagged data in Table \ref{tab:exp-pion-kaon}.
The total FF in this case is expressed up to the NLO as 
\begin{align}
\! \! \! 
{F^{h}}(z,Q^2)
& = {C_u} (z)  \otimes \! \bigg[ D_{u^+}(z,Q^2) 
+ \frac{1}{4}D_{d^+}(z,Q^2) 
+ \frac{1}{4}D_{s^+}(z,Q^2) \bigg] \!
+ {C_g} (z) \otimes D_g(z,Q^2) ,
\label{total FF-1}
\end{align}
where $C_u$ and $C_g$ denote the coefficient functions for 
$u$-quark and gluon, respectively. 
We note that the quark coefficient functions include
the charge-factor squared $e_q^{\, 2}$.
The gluon coefficient function is order $\alpha_s$ suppressed, 
so that the quark FFs are primarily determined 
from the $e^+e^-$ annihilation data. 
The evolution equations for those FFs are obtained 
from (\ref{eqn:evolution}) as
\begin{align}
\frac{\partial }{{\partial \ln Q^2}} D_{q_i ^+ } (z,Q^2)
& = \bar\alpha_s  P_{NS}^+(z) \otimes D_{q_i^ + } (z,Q^2) 
+ \bar\alpha_s {P_{PS}}(z) \otimes D_\Sigma (z,Q^2)
\nonumber \\[-0.2cm]
& \hspace{0.0cm}
+ 2 \, \bar\alpha_s  {P_{gq}}(z) \otimes D_g (z,Q^2),
\label{evolv-plus} 
\\
\frac{\partial }{{\partial \ln Q^2}} D_{g} (z,Q^2)
& = \bar\alpha_s {P_{qg}} (z) \otimes D_\Sigma (z,Q^2)
+ \bar\alpha_s  {P_{gg}} (z) \otimes D_g (z,Q^2),
\label{evolv-gluon}
\end{align}
where $\bar \alpha_s \equiv \alpha_s /(2 \pi)$
and $D_\Sigma  \equiv \sum_i^{N_f} D_{q_i^+}$.
Here, we have performed the standard decompositions 
for the splitting functions 
into flavor singlet ($S$) and nonsinglet ($V$)
terms \cite{esw-book} 
\begin{align}
P_{q_i q_j} = \delta_{ij} P_{qq}^V  + P_{qq}^{S} \, , \ \ 
P_{q_i \bar{q_j}} = \delta_{ij} P_{q\bar{q}}^V  + P_{q\bar{q}}^{S} \, ,
\end{align}
and we defined the nonsinglet ($NS$) and pure-singlet ($PS$) components 
of the $qq$-type splitting functions as 
\begin{align}
P_{q_i q_j} + P_{q_i \bar{q_j}} 
= \delta_{ij} \left(P_{qq}^V+P_{q\bar{q}}^V\right) 
 + P_{qq}^{S}+P_{q \bar q}^{S} 
\equiv \delta_{ij} P_{NS}^+  + {P_{PS}} \, .
\end{align}
Similarly, the scale evolution of the ``minus" component of the quark FF, 
$D_{q_i^-}\equiv D_{q_i}-D_{\bar{q}_i}$, is governed by 
\begin{align}
&\frac{\partial }{{\partial \ln Q^2}} D_{q_i ^- } (z,Q^2)
 = \bar\alpha_s  P_{NS}^-(z) \otimes D_{q_i^ - } (z,Q^2) \ ,
\label{evolv-minus}
\end{align} 
with the splitting function:
$P_{NS}^-\equiv P_{qq}^V - P_{q \bar{q}}^V$.

The solution of the evolution equation can be expressed concisely 
in terms of the evolution functions \cite{Furmanski:1981cw}.
For example, the solution of Eq.(\ref{evolv-minus}) is expressed as 
\begin{align}
D_{q_i^-} (z,Q^2) & = E_{NS}^{-} (z;\, Q^2, Q_0^2) 
\otimes D_{q_i^-} (z,Q_0^2) ,
\end{align}
where the evolution function $E_{NS}^-(z;Q^2,Q_0^2)$ is given 
in terms of the splitting functions $P_{NS}^-$ and QCD beta function. 
The solutions for the singlet FFs are also expressed in a matrix 
form with the evolution functions given by the singlet splitting functions.   
Substituting the solution for each flavor component, 
we rewrite the total FF as 
\begin{align}
F^{{h}}(z,{Q^2}) & = C_u (z) \otimes \bigg[ E_{NS}^+ (z;\, {Q^2},Q_0^2 ) 
\otimes \left\{ {\frac{1}{2}{D_{u^+}}(z,Q_0^2) 
- \frac{1}{4}{D_{d^+}}(z,Q_0^2)- \frac{1}{4}{D_{s^+}}(z,Q_0^2)} \right\}
\nonumber \\
& \ \ \ \ \ \ \ \ \ \ \ \ 
+\frac{1}{2}
\left\{ E_{\Sigma \Sigma } \left( z;\, {Q^2},Q_0^2 \right) 
\otimes D_{\Sigma}(z,Q_0^2)
+ E_{g \Sigma} \left( z;\, {Q^2},Q_0^2 \right) \otimes D_g(z,Q_0^2)
\right\} \bigg] 
\nonumber \\
& + {C_g} (z) \otimes \left[ {E_{q g}} (z;\, {Q^2},Q_0^2 ) 
\otimes {D_{\Sigma}(z,Q_0^2)}  
+ {E_{gg}}\left( z;\, {Q^2},Q_0^2 \right) \otimes D_g(z,Q_0^2) \right] ,
\label{total FF-2}
\end{align}
where $E_{\Sigma \Sigma }$, $E_{g \Sigma}$,
$E_{q g}$, and $E_{g g}$ are given by
the splitting functions $P_{NS}^+ +n_f P_{PS}$,
$2 n_f P_{gq}$, $P_{q g}$, and $P_{g g}$, respectively,
with the number of flavor $n_f$.
In Eq.(\ref{total FF-2}), three independent sets of FFs 
at the initial scale, namely, 
$D_{u^+}(z,Q_0^2)$, $D_{d^+}(z,Q_0^2)+D_{s^+}(z,Q_0^2)$ and $D_g(z,Q_0^2)$  
appear with different combinations of the evolution functions. 
The first two sets of FFs give the dominant contribution 
because of the relations: 
$C_u/e_u^2= \delta (1-z) +{\cal O}(\alpha_s)$,
$C_g= {\cal O}(\alpha_s)$;
$E_{NS}, E_{\Sigma\Sigma}, 
E_{gg}= \delta (1-z)+ {\cal O}(\alpha_s\ln(Q^2/Q_0^2))$;
$E_{q g},E_{g\Sigma}= {\cal O}(\alpha_s \ln(Q^2/Q_0^2))$. 
The relative weights of these three sets change 
as the scale $Q^2$ evolves, so that these sets can be separated 
from each other in principle. 
However, if we only rely on the pure-QED process
with the intermediate $\gamma$,
a clear separation of these sets requires high precision data 
at least at three distinct energies.        

In the case of the $e^+e^-$ annihilation at the Z pole, 
the total FF is given by 
\begin{align}
& \! \! \! 
F^{h} (z,M_Z^2) = {{\tilde C}_q} (z) \otimes 
\bigg[ (c_{V}^{u \, 2} + c_{A}^{u \, 2})
\big\{ D_{u^ + } (z,M_Z^{\, 2}) + D_{c^ + } (z,M_Z^{\, 2}) \big\} 
\nonumber \\[-0.3cm]
&  \ \ \ \ \ 
+ (c_{V}^{d \, 2} + c_{A}^{d \, 2})
\big\{ D_{d^ + } (z,M_Z^{\, 2}) + D_{s^ + } (z,M_Z^{\, 2}) 
      + D_{b^ + } (z,M_Z^{\, 2}) \big\} \bigg] 
+ {C_g} (z) \otimes D_g(z,M_Z^{\, 2}) .
\label{total FF-3}
\end{align}
Here, $\tilde C_q$ is given by replacing the quark-charge square $e_q^2$
within $C_q$ of Eq. (\ref{total FF-2}) by weak couplings and then taking
the couplings $c_{V}^{q \, 2} + c_{A}^{q \, 2}$ out from the coefficient 
function.
The vector and axial-vector couplings are given by 
$c_V^q=T_q^3 -2 e_q \sin^2 \theta_W$ and $c_A^q=T_q^3$, respectively
\cite{hm-book}. 
Taking $\sin^2 \theta_W=0.23$ \cite{Beringer:1900zz}, we have 
$c_{V}^{u \, 2} + c_{A}^{u \, 2}=0.287$ and 
$c_{V}^{d \, 2} + c_{A}^{d \, 2}=0.370$, which are roughly the same. 
Therefore, we have
\begin{align}
\! \! \! \! \! \! \! \! \! 
F^{h} (z,M_Z^2)
& \approx {\tilde C}_q^\prime (z) \otimes D_\Sigma (z,M_Z^{\, 2}) 
+ {C}_g (z) \otimes D_g(z,M_Z^{\, 2}) 
\nonumber \\
& 
= {\tilde C}_q^\prime (z) \otimes 
\left[ \, E_{\Sigma\Sigma}(z; M_Z^2,Q_0^2) \otimes D_\Sigma (z,Q_0^2) \right.
\left.
+E_{g \Sigma}(z; M_Z^2,Q_0^2)\otimes D_g (z,Q_0^2) \, \right]
\nonumber\\
&   
+ {C}_g (z) \otimes 
\left[ \, E_{q g}(z; M_Z^2,Q_0^2) \otimes D_\Sigma (z,Q_0^2) \right.
\left.
+E_{g g}(z; M_Z^2,Q_0^2)\otimes D_g (z,Q_0^2) \, \right] ,
\label{total FF-4} 
\end{align} 
where ${\tilde C}_q^\prime = 0.33 \, {\tilde C}_q$, 
which is approximately the average of 
$0.287 \, {\tilde C}_q$ and $0.370 \, {\tilde C}_q$,
and the dominant contribution comes only from the singlet quark FF.

Now, we discuss separation of the quark flavors and determination 
of the gluon FF by using the $e^+e^-$ annihilation data both 
at and below the Z-pole.    
Note that in Eqs. (\ref{total FF-2}) and (\ref{total FF-4}),
(or (\ref{total FF-3}) to be precise), $D_{u^+}$ and 
$D_{d^+}+D_{s^+}$ appear as the dominant contributions
with different relative weights, due to the difference of 
the electric and weak charges of quarks. Therefore, even if we ignore 
the evolution effects, those two sets of FFs can be 
separately determined in a relatively clear manner.      
The impact of the B-factory data is understood as follows. 
\begin{itemize}
\item[(1)]
In the global fit before the B-factory data, the $e^+e^-$ annihilation 
data have been dominated by the LEP and SLD data, from which 
the singlet FF is mainly determined.
\item[(2)]
Then, by combining with the pure-QED process data at lower energies 
with a moderate precision, the two components of the quark FFs,
$D_{u^+}(Q_0^2)$ and $D_{d^+}(Q_0^2)+D_{s^+}(Q_0^2)$ can be separated 
as mentioned above.  
\item[(3)]
On the other hand, determination of the gluon FF, especially  
its separation from the singlet FF must rely on changes of their relative 
weights through the scale evolution, so that the precise determination 
of the gluon FF requires high precision data at energy scales 
much lower than the Z mass. The B-factory data 
are considered to provide such information.
\end{itemize}

So far, we have discussed flavor separation and determination of 
the gluon FF based on the cross section formula and the scaling violation. 
Now, we introduce the favored and disfavored FFs ($D_{fav}$, $D_{dis}$)
at the initial scale. 
Let us take the pion case, where we have the initial relations at $Q_0^2$:
\begin{align}
\! \! \! 
D_{u^ \pm }^{\pi^+}(z,Q_0^2) =  \pm D_{d^ \pm }^{\pi^+}(z,Q_0^2), \ 
D_{s^ - }^{\pi^+}(z,Q_0^2) = 0, \ 
D_{u^ + }^{\pi^+}(z,Q_0^2) - D_{u^ - }^{\pi^+}(z,Q_0^2) 
= D_{s^ + }^{\pi^+}(z,Q_0^2) , 
\label{pi-relation}
\end{align}
which are equivalent to Eqs. (\ref{favored}) and (\ref{disfavored}).
From Eqs. (\ref{evolv-plus}) and (\ref{evolv-minus}),   
one can see that the first two relations are preserved, whereas
the third relation is violated by the scale evolution
to $Q^2 \ne Q_0^2$.
This assignment of favored and disfavored FFs assumes the constituent 
quark model, the OZI (Okubo-Zweig-Iizuka) rule, 
and the flavor $SU(3)$ symmetry in fragmentation process 
at the initial scale $Q_0^2$. 
In the actual fitting procedure, this assignment is nothing but 
a change of the independent sets of the initial FFs: 
\begin{align}
D_{u^+}^{\pi^+}(z,Q_0^2) &\ra D_{fav}^{\pi^+}(z,Q_0^2)
+D_{dis}^{\pi^+}(z,Q_0^2) , 
\nonumber\\
D_{d^+}^{\pi^+}(z,Q_0^2)+D_{s^+}^{\pi^+}(z,Q_0^2)
&\ra D_{fav}^{\pi^+}(z,Q_0^2)
+3 D_{dis}^{\pi^+}(z,Q_0^2) ,
\end{align}
in Eqs. (\ref{total FF-2}) and (\ref{total FF-3}).
Therefore, 
$D_{fav}^{\pi^+}$ and $D_{dis}^{\pi^+}$
can be well-determined separately 
by using the LEP and SLD data and the pure-QED process data
as discussed above.      
Instead, one can also use a more general parametrization 
without assuming the third relation in Eq. (\ref{pi-relation}), 
by taking 
$D_{u,\bar{d}}(Q_0^2) = D_{fav}^{\pi^+}(Q_0^2)$, 
$D_{\bar{u},d}(Q_0^2) = D_{dis(1)}^{\pi^+}(Q_0^2)$ and 
$D_{s,\bar{s}}^{\pi^+}(Q_0^2)=\tilde{D}_{dis(2)}^{\pi^+}(Q_0^2)$.
Advantages of this parametrization are that it takes the $SU(3)$ breaking 
effect in the fragmentation process and is compatible with the scale
evolution.  
However, since the separation of these three sets of FFs must rely on 
the scale evolution, it is as hard as the determination of the gluon FF. 

For the kaon, we have the initial relations: 
\begin{align}
D_{d^ - }^{K^+}(z,Q_0^2) & = 0 \ , 
\nonumber \\
D_{d^ + }^{K^+}(z,Q_0^2) & = 
D_{u^ + }^{K^+}(z,Q_0^2)- D_{u^ - }^{K^+}(z,Q_0^2) 
= D_{s^ + }^{K^+}(z,Q_0^2)- D_{s^ - }^{K^+}(z,Q_0^2) \ ,
\label{K-relation}
\end{align}
where the first relation is preserved, while the second and third 
ones are violated by the scale evolution.
The flavor separation in the kaon case is similar to the one in the case 
we have just discussed above as a generalized parametrization for pion.
It is not easy to determine two favored functions $D_{u}^{K^+}$ 
and $D_{\bar s}^{K^+}$, one disfavored function $D_{dis}^{K^+}$, 
and the gluon function $D_{g}^{K^+}$ simultaneously 
in comparison with the one favored function for the pion.

To summarize, flavor separation of two sets of quark FFs can be 
done clearly by using the $e^+e^-$ annihilation at and below 
the $Z$-pole, which is due to the difference of the quark coupling 
to the virtual photon and $Z$. 
On the other hand, flavor separation of three sets of quark FFs 
and determination of the gluon FF requires the effects of 
the scaling evolution. The data from B-factory measurements, obtained
at an energy much lower than the $Z$ mass with a wide $z$ range, 
are expected to provide a major assistance for 
the latter analyses. Specifically, the uncertainty of the gluon FF 
is expected to decrease significantly once the B-factory data are included 
in the analysis and provided that the quark FFs have been determined 
with enough accuracy without the B-factory data.

\subsection{$\chi^2$ analysis}
\label{chi2-analysis}

As shown in Fig. \ref{fig:zq2}, the experimental data are taken
at various $Q^2$ values, which are different from the initial
scale $Q_0^2=1$ GeV$^2$.
The timelike DGLAP equation (\ref{eqn:evolution}) 
is used for calculating the $Q^2$ evolution.
The scale parameter is taken 
$\Lambda^{(4)}_{NLO}$=0.323 GeV for four flavors \cite{th-mass}.
This value is changed with the number of flavors 
at the heavy-quark thresholds, and it is $\alpha_s (M_z)=0.119$ 
at the Z mass. The charm and bottom quark masses are $m_c=1.43$ GeV 
and $m_b=4.3$ GeV in our analyses.
These values are the same in the previous analysis \cite{hkns07}, and
the $\overline {\rm MS}$ scheme is used in the NLO. 
The numerical solution is calculated by the method of Ref. \cite{hk-q2evol}
for the $Q^2$ evolution. The initial functions are evolved to the experimental
points of $Q^2$. Then, the theoretical fragmentation function $F_{j}^{theo}$ is
calculated by Eq. (\ref{eqn:def-ffqqbarg}) 
at the experimental $Q^2$ point 
to compare it with the experimental one $F_{j}^{data}$ in 
Eq. (\ref{eqn:def-ff}).

The total $\chi^2$ is then given by \cite{chi2-ref}
\begin{equation}
\chi^2 = \sum_{i=1}^n \left[ 
               \sum_{j=1}^m \frac{(D_j - T_j/N_i)^2}
                {(\sigma^{\text{exp}}_{j})^2}
                + \frac{(N_i-1)^2}{\sigma_{N_i}^2} \right]
\label{eqn:chi2}
\end{equation}
where $T_j$ and $D_j$ are theoretical calculation and a datum, respectively,
and $N_i$ ($\sigma_{N_i}$) is the normalization factor (its error)
for the data set $i$. 
For simplification of discussions about consistency of the B-factory 
data sets with other measurements, the overall normalization factors 
are introduced. Correlations of the systematic errors are not considered, 
and the experimental errors are calculated from systematic 
and statistical errors by
$(\sigma_j^{\text{exp}})^2 
= (\sigma_j^{\text{stat}})^2 + (\sigma_j^{\text{sys}})^2 $.
As the overall scale shifts on the data, the free factors are used only 
for the B-factory data sets. Its values for the other data sets are fixed
as $N_i=1$.
The optimum parameters are obtained 
by minimizing $\chi^2$ by the CERN subroutine {\tt MINUIT} \cite{minuit}.
Uncertainties of the PDFs have been calculated, for example, as
explained in Refs. \cite{pdf-errors,errors}.
Here, the Hessian method is employed to calculate the uncertainties
of the FFs . 

We denote the parameters of the FFs as $\xi_i$ 
($i$=1, 2, $\cdot \cdot \cdot$, $N$).
The total $\chi^2$ is expanded around the minimum point $\hat \xi$
by keeping only the quadratic term:
\begin{equation}
 \Delta \chi^2 (\xi) = \chi^2(\hat{\xi}+\delta \xi)-\chi^2(\hat{\xi})
        =\sum_{j,k} H_{jk} \, \delta \xi_j \, \delta \xi_k \ ,
\label{eq:chi2expand}
\end{equation}
where $H_{jk}$ is called Hessian. The errors of the FFs are estimated
by supplying the value of $\Delta \chi^2$, which determines 
the confidence level. Using the Hessian matrix, which is numerically
obtained by running the {\tt MINUIT} code, we calculated the errors
of the FFs by 
\begin{equation}
[\delta D_i^h (z)]^2=\Delta \chi^2 \sum_{j,k}
\left( \frac{\partial D_i^h (z,\xi)}{\partial \xi_j}  \right)_{\hat\xi}
H_{jk}^{-1}
\left( \frac{\partial D_i^h (z,\xi)}{\partial \xi_k}  \right)_{\hat\xi}
\, .
\label{eqn:ddih}
\end{equation}
We note that the Hessian method assumes that the quadratic expansion
of Eq. (\ref{eq:chi2expand}) would give a good approximation of $\chi^2$ 
in the vicinity of the minimum. 
A detailed study employing the Lagrange multiplier technique 
suggests that this assumption is indeed feasible in most cases 
\cite{Epele:2012vg}.

As well known, $\Delta \chi^2=1$ gives the confidence level of 68\%,
namely the one-$\sigma$-error range, if the number of parameter is one.
For the multiparameters with $N$ degrees of freedom, the confidence
level could be calculated by
\begin{equation}
        P=\int_0^{\Delta \chi^2} \frac{1}{2\ \Gamma(N/2)} 
        \left(\frac{S}{2}\right)^{\frac{N}{2}-1} 
               \exp\left(-\frac{S}{2} \right) dS \ ,
\label{eq:dchi2}
\end{equation}
as explained in Ref. \cite{del-chi2}. 
As a $\Delta \chi^2$ criterion, we applied this one-$\sigma$ range
for the probability distribution in the multi-parameter space.
We use the $\Delta \chi^2$ value to obtain the one-$\sigma$-error range
$P=0.6826$ by Eq. (\ref{eq:dchi2}).
It is given by $\Delta \chi^2=15.94$ for $N=14$ in the pion analysis
and $\Delta \chi^2=19.20$ for $N=17$ in the kaon.
Roughly, these values correspond to $\Delta \chi^2 \sim N$.
One may note that if a physical value is calculated by the determined 
fragmentation functions with multi-parameters, it could be
statistically right to show the one-$\sigma$ range by $\Delta \chi^2=1$. 
Our choice of $\Delta \chi^2 \sim N$ could be considered 
as a tolerance value in the unpolarized PDF analysis.
In the PDF studies, different values are used for
$\Delta \chi^2$ depending on analysis groups.
More details are explained in 
Ref. \cite{errors} for the polarized PDFs,
the interested reader may look at this article.

\section{\label{results} Results}


\begin{table}[t]
\caption{Each $\chi^2$ contribution in the pion and kaon NLO analysis.
The normalization factors are also shown for Belle and BaBar.
The values in the parentheses (\ \ ) indicate
the ones of the HKNS07 parametrization \cite{hkns07} without
the Belle and BaBar data.}
\label{tab:chi2-pion-kaon}
\centering
\begin{tabular}
{l@{\extracolsep{0ptplus1fil}}c@{\extracolsep{0ptplus1fil}}c
@{\extracolsep{0ptplus1fil}}c@{\extracolsep{20ptplus1fil}}c
@{\extracolsep{0ptplus1fil}}c@{\extracolsep{0ptplus1fil}}c}
experiment           & \# of $\pi$ data  & $\chi^2 (\pi)$  & $N_i(\pi)$
                     & \# of K data  & $\chi^2 (\text{K})$ & $N_i(\text{K})$ \\
\hline
Belle                &    78   &  44.3 ( $-$ ) &  0.94 \ & 78   &  18.8 ( $-$ )   & 1.03 \\
BaBar                &    36   &  141.6 ( $-$ ) \, &  0.90 \ & 36  &  37.8 ( $-$ ) & 0.96 \\
TASSO                &    29   &  55.2 (51.9)     & $-$ & 18    &  26.2 (25.0)     & $-$ \\
TPC                  &    18   &  14.2 (27.3)     & $-$ & 17    &  40.5 (15.2)     & $-$ \\
HRS                  &  \, 2   &  \, 4.1 \, (2.0) & $-$ & \, 3  &  \, 0.4 \, (0.4) & $-$ \\
TOPAZ                &  \, 4   &  \, 3.9 \, (2.6) & $-$ & \, 3  &  \, 1.7 \, (0.8) & $-$ \\
SLD (all)            &    29   &  41.7 (10.6)     & $-$ &  29   &  17.0 (12.3)     & $-$ \\
SLD (u,d,s)          &    29   &  90.7 (36.4)     & $-$ &  29   &  59.3 (57.2)     & $-$ \\
SLD (c)              &    29   &  36.0 (26.1)     & $-$ &  29   &  39.3 (32.4)     & $-$ \\
SLD (b)              &    29   &  67.7 (66.4)     & $-$ &  28   &  99.0 (88.7)     & $-$ \\
ALEPH                &    22   &  47.5 (24.0)     & $-$ &  18   &  \, 8.7 (12.8)   & $-$ \\
OPAL                 &    22   &  57.1 (45.8)     & $-$ &  10   &  \, 9.1 (11.5)   & $-$ \\
DELPHI (all)         &    17   &  37.2 (48.6)     & $-$ &  27   &  15.0 (15.2)     & $-$ \\
DELPHI (u,d,s)       &    17   &  24.8 (31.1)     & $-$ &  17   &  23.0 (22.1)     & $-$ \\
DELPHI (b)           &    17   &  64.4 (60.8)     & $-$ &  17   &  10.1 (11.7)     & $-$ \\
\hline
total        & 378 (264) \, & 730.4 (433.5) \, \,  & & 359 (245) \, & 405.9 (305.1) \, \, & \\
total/d.o.f. &              &   2.02 (1.73)        & &              & 1.19 (1.34)  &     \\
\end{tabular}
\end{table}

\begin{figure}[t]
\vspace{0.3cm}
\hspace{0.5cm}
\includegraphics[width=0.80\textwidth]{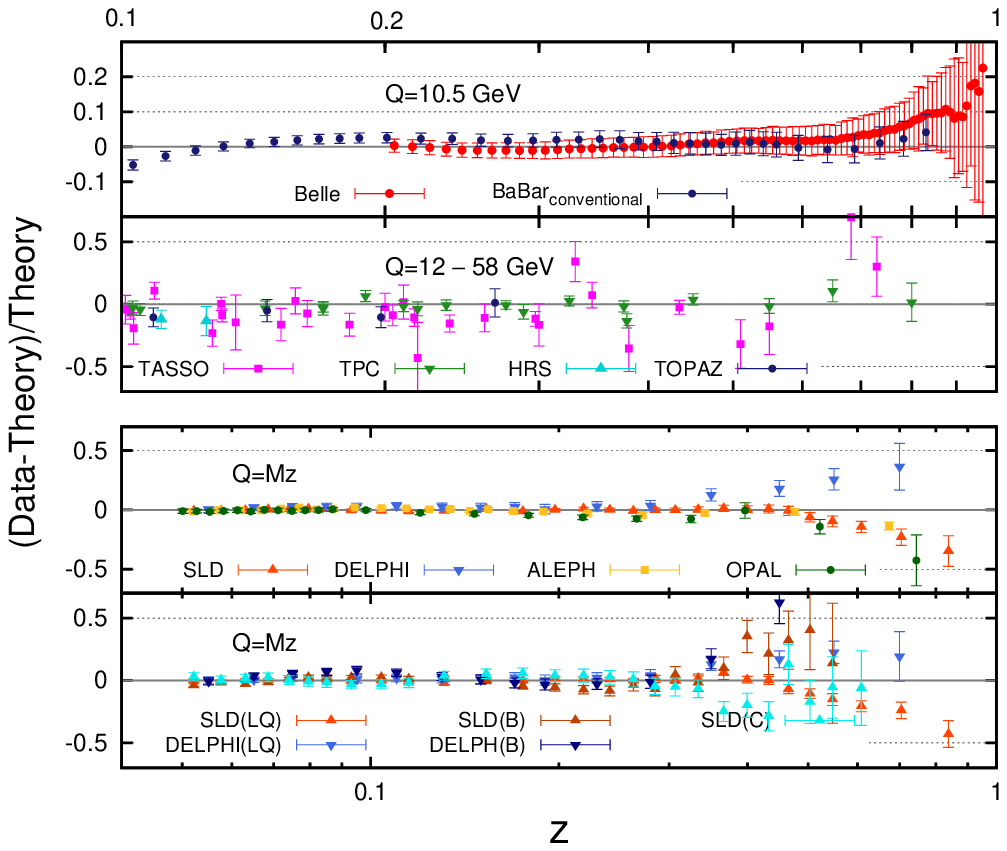}
\vspace{-0.3cm}
\caption{Comparison of theoretical results at the NLO with 
charged-pion production data.
The $z$ scale ranges from 0.1 to 1 in the upper two figures 
(Belle, BaBar; TASSO, TPC, HRS, TOPAZ), and it is 
from 0.04 to 1 in the lower two figures 
(SLD, DELPHI, ALEPH, OPAL; SLD(LQ,B,C), DELPHI(LQ,B)).}
\label{fig:pion-data-nlo}
\end{figure}

Analysis results are shown by $\chi^2$ values for all the used experimental
data sets of the pion and kaon in Table \ref{tab:chi2-pion-kaon}.
Here, the values in the parentheses are $\chi^2$ values of
the HKNS07 NLO analysis \cite{hkns07} without the B-factory data.
In the pion analyses, the Belle and BaBar $\chi^2$ values in the table
consist of the first (data) and second (normalization) 
terms in Eq. (\ref{eqn:chi2}), and they are 
44.3=24.1\,(1st)+20.2\,(2nd) for the Belle and
141.6=44.9\,(1st)+96.7\,(2nd) for the BaBar. 
We notice that the normalization contributions to the total $\chi^2$ 
are large. The first $\chi^2$ value of the Belle is very small by 
considering the number of data of 78, whereas the agreement with 
the BaBar data is marginal. 
However, the normalization shifts are significantly larger
than experimental estimations for the overall normalization
errors, 1.4\% (0.98\%) for Belle (BaBar).
It indicates that the B-factory measurements are not completely
consistent with the previous data.
We notice in the table that the $\chi^2$ values become larger than
the previous HKNS07 ones for the SLD measurements except 
for the bottom-quark data,
whereas the $\chi^2$ values stay almost the same for the TASSO, OPAL, 
and DELPHI. 

Actual comparisons of the NLO pion results with the data are shown 
in Fig. \ref{fig:pion-data-nlo}, where the fractional differences, 
(Data$-$Theory)/Theory, are shown for all the data sets.
As suggested by the small $\chi^2$ for the Belle data in 
Table \ref{tab:chi2-pion-kaon}, the agreement with the Belle data is 
very good in the top figure ($Q=10.5$ GeV) of 
Fig. \ref{fig:pion-data-nlo}. 
It is also clear that the Belle and BaBar
data have excellent accuracies in comparison with the other ones.
Because of the tiny errors of the B-factory data, the parametrized
functions converge so as to fit these data as shown in the top figure
of Fig. \ref{fig:pion-data-nlo}.
It results into deviations from some data sets. 
There are noticeable differences from the SLD data
although their errors are small, which makes large $\chi^2$ contributions
as noticed in Table \ref{tab:chi2-pion-kaon} and suggests that
there are some discrepancies between the B-factories and SLD ones.
On the other hand, the LEP (OPAL, DELPHI) data have good agreement
with the obtained theoretical functions although there are slight
differences from the LEP-ALEPH data. 

\begin{figure}[t!]
\vspace{0.0cm}
\hspace{-1.1cm}
\includegraphics[width=0.66\textwidth]{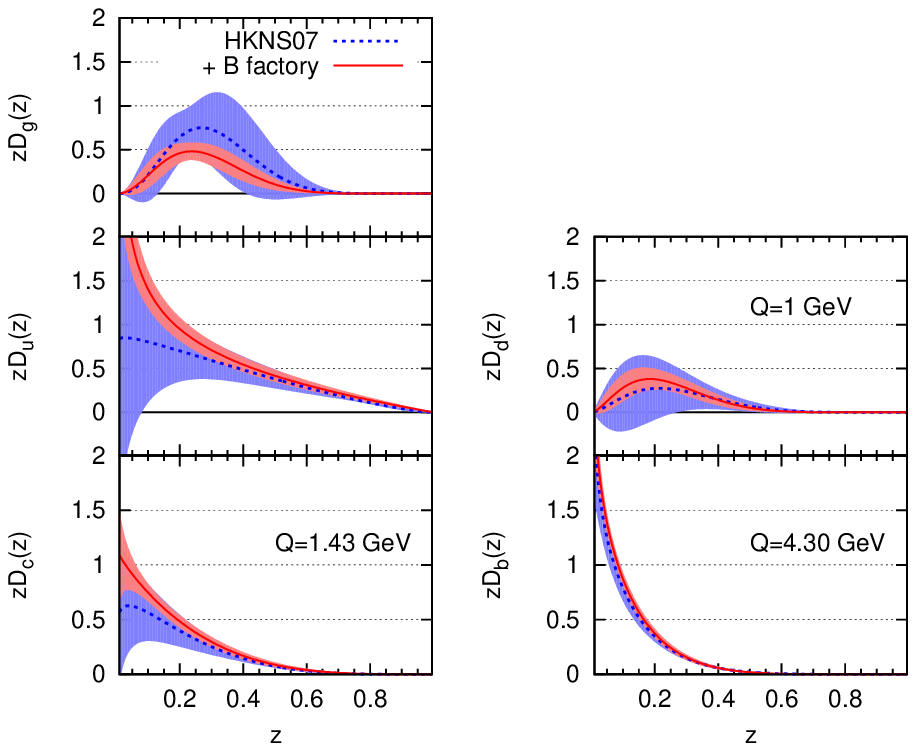}
\hspace{-2.5cm}
\includegraphics[width=0.66\textwidth]{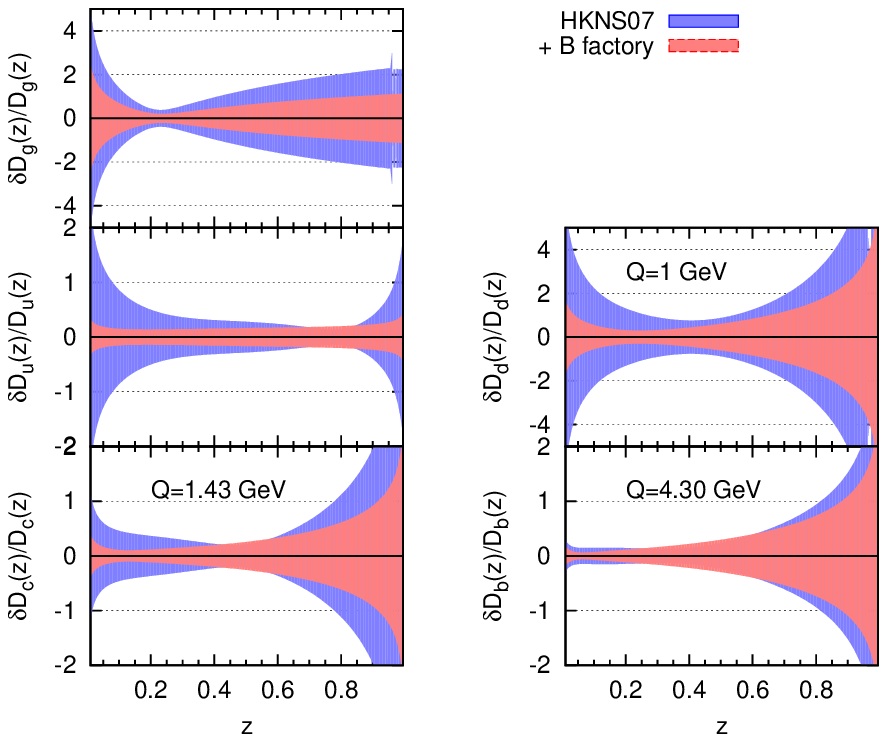}
\vspace{-1.7cm}
\hspace{3.2cm}{$(a)$}
\hspace{7.3cm}{$(b)$}
\vspace{+1.5cm}
\caption{$(a)$ Fragmentation functions and their uncertainties
          are shown for $\pi^+$ at $Q^2$=1 GeV$^2$, $m_c^2$, and $m_b^2$
          in the NLO analysis.
         $(b)$ Relative uncertainties are shown.
          The dashed and solid curves indicate HKNS07 and current (HKKS16) 
          results, and the HKNS07 and HKKS16 uncertainties are shown by 
          the dark- and light-shaded bands, respectively.}
\label{fig:pion-ff-q-1-nlo}
\end{figure}

The determined FFs of the pion ($\pi^+$) are shown for the NLO
in Fig. \ref{fig:pion-ff-q-1-nlo}.
In order to show the improvements due to the B-factory data,
the FFs and their uncertainties of the previous
version HKNS07 are shown for comparison in Fig. \ref{fig:pion-ff-q-1-nlo} $(a)$.
We find that the gluon and light-quark FFs are changed 
from the HKNS07 ones.
Especially, the favored fragmentation function $D_u^{\pi^+}$
increases steeply at small $z$ as shown in this figure,
and it is the cause of the $\chi^2$ increase for SLD.
Moreover, the second moment of $D_u^{\pi^+}$ becomes larger
than the HKNS07 one, and it reaches almost the upper limit of
the momentum sum for the current analysis.
It could suggest that a more flexible functional form is required 
for the favored FF in order to keep the sum rule.
However, even if the flexible form is introduced, it seems to be difficult
to explain the SLD data because they cannot be fit at the same time with
the B-factory and DELPHI data because of their inconsistencies
in Fig.\,\ref{fig:pion-data-nlo}. 

There are significant reductions of the uncertainty bands in all the FFs. 
There are two sources of these reductions due to the B-factory data.
First, the precise data at $\sqrt{s} \simeq 10.5$ GeV combined with 
other accurate data taken at larger $Q^2$, mainly at $Q^2=M_Z^2$, 
make it possible to determine the quark FFs precisely.
It is apparently shown by the smaller errors in the favored 
and disfavored quark FFs.
This is due to the fact that the quark-pair production is dominated 
by the electro- and weak-interaction in the B-factory and LEP/SLD data, 
respectively, so that $F^{\,\pi^\pm}$ is given by $D_{fav}$ and $D_{dis}$ 
with different relative weights 
in Eqs. (\ref{total FF-2}) and (\ref{total FF-3}). 
Second,  the scaling violation information
between the B-factory and $Q^2=M_Z^2$ data
should impose a constraint on the gluon FF.
In fact, the gluon fragmentation function is determined
more accurately by including the B-factory data.
Furthermore, if the gluon function is obtained more accurately,
it affects the better determination of the quark 
FFs due to error correlation effects.
Thus, these results for the pion FFs are consistent with  
the discussion given in Sec. \ref{flavor}.

\begin{figure}[t]
\vspace{0.3cm}
\hspace{0.5cm}
\includegraphics[width=0.80\textwidth]{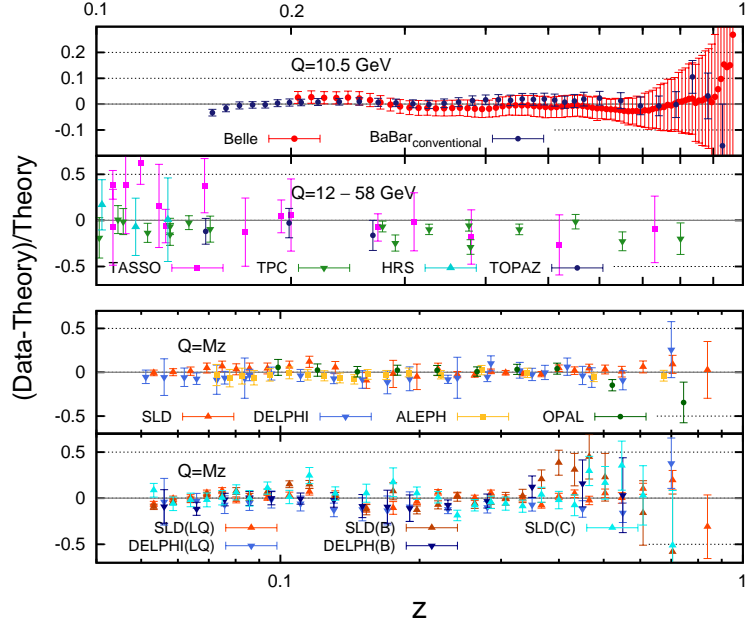}
\vspace{-0.3cm}
\caption{Comparison of theoretical results at the NLO with 
charged-kaon production data.
The $z$ scale ranges from 0.1 to 1 in the upper two figures 
(Belle, BaBar; TASSO, TPC, HRS, TOPAZ), and it is 
from 0.04 to 1 in the lower two figures 
(SLD, DELPHI, ALEPH, OPAL; SLD(LQ,B,C), DELPHI(LQ,B)).}
\label{fig:kaon-data-nlo}
\end{figure}

Kaon data are also analyzed by including the new B-factory data.
First, the $\chi^2$ values are shown in Table \ref{tab:chi2-pion-kaon}.
In the pion case, there are some differences in the $\chi^2$ values
of the HKNS07 data sets with or without the B-factory ones; 
however, the $\chi^2$ values stay almost the same as the HKNS07 ones 
for all the kaon experimental sets except for TPC 
even if the B-factory data are included.
Furthermore, the normalization shifts are not as large as the pion factors,
and they are 1.03 and 0.96 for Belle and BaBar, respectively, 
whereas the normalization
errors are 1.4\% (0.98\%) in Belle (BaBar). It indicates that 
the B-factory kaon data are nearly consistent with previous data, although
the normalizations should be still introduced for obtaining sensible results.
The first (data) and second (normalization) contributions
to $\chi^2$ in Eq. (\ref{eqn:chi2}) are 
18.8=15.0\,(1st)+3.8\,(2nd) for the Belle and
37.8=21.6\,(1st)+16.1\,(2nd) for the BaBar. 
It is noteworthy that the first $\chi^2$ values 
for the B-factories are very small in the same way 
with the pion analysis. The data are valuable for determining the kaon FFs.
The second normalization contributions to $\chi^2$ are much smaller
than the ones for the pion.
We show the comparison of the obtained kaon FFs with the data
in Fig. \ref{fig:kaon-data-nlo}. 
We applied the $z$ cut ($z>0.15$) for the BaBar as explained 
in Sec. \ref{data}.
In general, the figure shows a good agreement with
almost all the data sets within their errors as also indicated
by the $\chi^2$ values in Table \ref{tab:chi2-pion-kaon}.

\begin{figure}[h!]
\hspace{-1.1cm}
\includegraphics[width=0.66\textwidth]{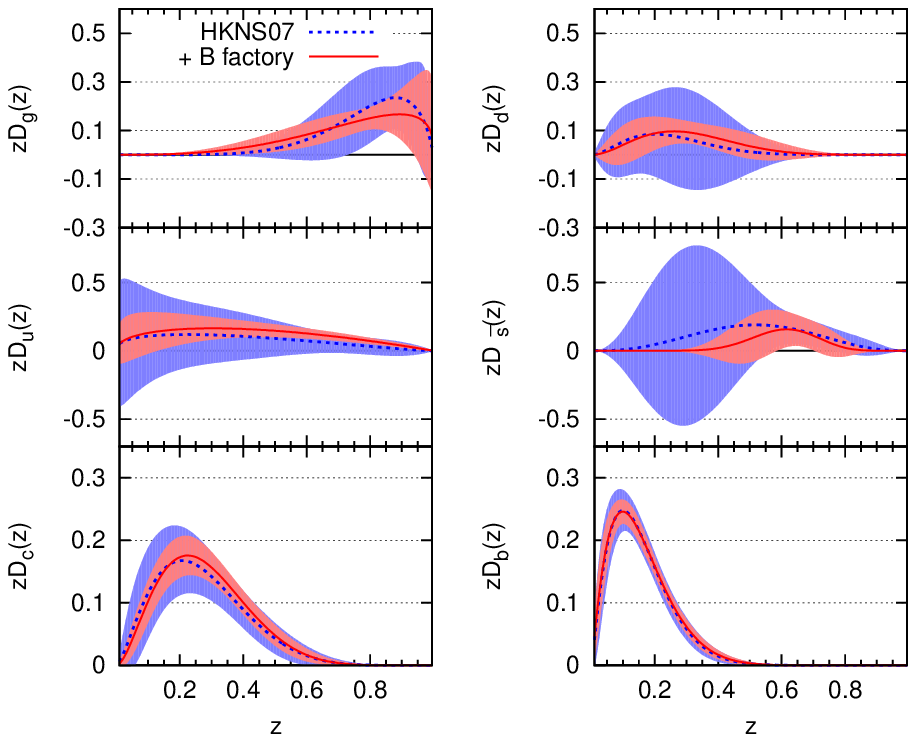}
\hspace{-2.5cm}
\includegraphics[width=0.66\textwidth]{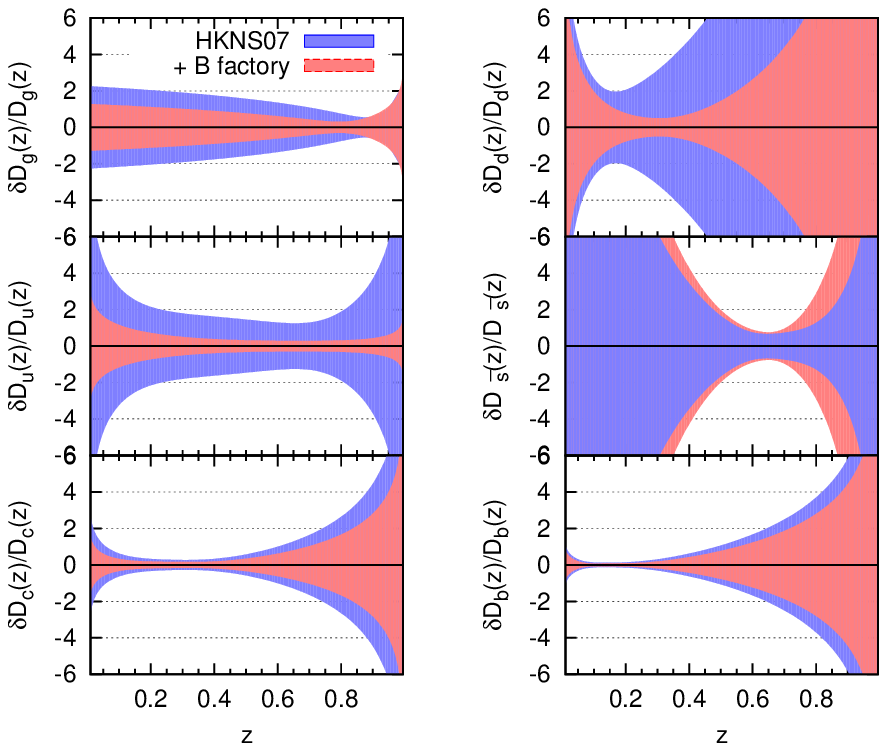}
\vspace{-1.7cm}
\hspace{3.2cm}{$(a)$}
\hspace{7.3cm}{$(b)$}
\vspace{+1.5cm}
\caption{$(a)$ Fragmentation functions and their uncertainties
          are shown for $K^+$  at $Q^2$=1 GeV$^2$, $m_c^2$, and $m_b^2$
          in the NLO analysis.
         $(b)$ Relative uncertainties are shown.
          The dashed and solid curves indicate HKNS07 and current (HKKS16) 
          results, and the HKNS07 and HKKS16 uncertainties are shown by 
          the dark- and light-shaded bands, respectively.}
\label{fig:kaon-ff-q-1-nlo}
\end{figure}

In order to illustrate the impact of the B-factory data on
the determination of the kaon FFs, the obtained functions
are compared with the HKNS07 ones with uncertainties in
Fig. \ref{fig:kaon-ff-q-1-nlo} $(a)$, 
and relative uncertainties are shown in $(b)$.
There are variations from the previous analysis HKNS07 
in the favored functions ($D_u^{K^+}$, $D_{\bar s}^{K^+}$).
In this analysis, the second moment of the favored one $D_{\bar s}^{K^+}$ 
becomes smaller than the HKNS07 one and it is comparable to
the one of the disfavored function $D_d^{K^+}$; however,
the moment of $D_{\bar s}^{K^+}$ is still larger than
the one of $D_d^{K^+}$.
We notice in Figs. \ref{fig:pion-ff-q-1-nlo} and \ref{fig:kaon-ff-q-1-nlo}
that the gluon and heavy-quark ($c$ and $b$) fragmentation functions of 
the kaon are shifted toward larger-$z$ regions. It may be interpreted
in the following way. In order to create $\pi^+$ from $g$, $c$, or $b$,
two quark-pair ($u\bar u$ and $d\bar d$) creations should occur, whereas
they are $u\bar u$ and $s\bar s$ for $K^+$. Since the $s$-quark is heavier than
$d$-quark, the initial parton ($g$, $c$, or $b$) should have more energy
to fragment into $K^+$. It means that the function $D_{g,c,b}^{K^+} (z)$ 
is distributed at relatively larger $z$ than $D_{g,c,b}^{\pi^+} (z)$.

Figures \ref{fig:kaon-ff-q-1-nlo} $(a)$ and $(b)$
indicate that the B-factory data
significantly improve the determination of the kaon FFs,
especially the light-quark and gluon FFs. 
In the kaon case, we considered two favored functions
instead of one favored function in the pion. 
There are four types: two favored, one disfavored,
and gluon functions, in addition to the charm and bottom functions.
Considering the current experimental situation, namely accurate
data exist mainly at $\sqrt{s}\simeq$10.5 GeV and $M_Z$,
the four independent functions could be considered as redundant
for determining all of them simultaneously.
It is the reason why the error bands are rather large
in comparison with the pion errors.
Nonetheless, it is good to obtain more accurate kaon FFs
due to the B-factory measurements because the determined FFs
are used in kaon-production processes, for example, in probing
the strange-quark distribution in the nucleon \cite{n-strange}
and strange-quark contribution to the nucleon spin content
\cite{sdis-delta-s}.
Furthermore, accurate measurements of FFs for exotic hadron
candidates, such as $f_0$(975), $a_0$(975), and $\Lambda$(1405), 
could provide valuable information on their internal
structure through the favored and disfavored FFs \cite{hkos08}.

\section{Summary}\label{summary}

We analyzed experimental data on hadron productions in electron-positron
annihilation, $e^++e^- \to h+X$, for determining fragmentation
functions. A recent experimental development is that the Belle and BaBar 
collaborations published high-statistics data on charged pion and kaon 
productions. 
In this article, we reported our results on the fragmentation functions 
of the pion and kaon with uncertainties estimated by the Hessian method
in order to show impacts on the B-factory data on the FF determination.
The uncertainties are compared between the analysis results 
and the previous HKNS07 ones without the B-factory information.

It was shown that the B-factory measurements contribute 
to significant reductions of the uncertainties, especially 
in the gluon and light-quark fragmentation functions. 
The B-factory data are taken at $\sqrt{s}\simeq 10.5$ GeV
and other accurate data are obtained at 91.2 GeV, so that the scaling
violation of the fragmentation functions became clear for the first time.
Furthermore, the electric and weak charges are different depending
on the quark type, a light-quark flavor separation also became possible. 
These two factors contributed to a more accurate determination 
of the gluon and light-quark functions. However, we noticed
some tensions between the Belle and BaBar data sets and also
between the B-factory and other data sets.
The more precise fragmentation functions, by using the Belle and
BaBar measurements, could make it possible for finding new physics 
and the details of nucleon structure at high-energy hadron facilities. 

\section*{Acknowledgements}
The authors thank Martin Leitgab, Matthias Grosse Perdekamp, 
and Ralf Seidl for communications on Belle measurements,
and they thank David Muller for explanations on BaBar experiments.
This work was partially supported
by JSPS KAKENHI Grant Numbers, JP21105006, JP25105010, and JP15K05061.

\vspace{-0.0cm}



\begin{thebibliography}{00}
\bibitem{hkns07} M. Hirai, S. Kumano, T.-H. Nagai, and K. Sudoh,
                         Phys. Rev. D {\bf 75}, 094009 (2007).
\bibitem{high-pt} F. Arleo, D. d'Enterria, and A. S. Yoon, 
                       JHEP {\bf 06}, 035 (2010);
      B. Abelev {\it et al.}  (ALICE Collaboration),
             Phys. Lett. B {\bf 717}, 162 (2012).
\bibitem{ffs_before_2006}     
                  B. A. Kniehl, G. Kramer, and B. P\"otter,
                     Nucl. Phys. B {\bf 582}, 514 (2000);
                  S. Kretzer, Phys. Rev. D {\bf 62}, 054001 (2000);
                  S. Albino, B. A. Kniehl, and G. Kramer,
                     Nucl. Phys. B {\bf 725}, 181 (2005).
\bibitem{ffs-summary} S. Albino {\it et al.}, arXiv:0804.2021 [hep-ph];
                      F. Arleo, Eur. Phys. J. C {\bf 61}, 603 (2009);
                      F. Arleo and J. Guillet, 
                          online generator of FFs at
                          http://lapth.cnrs.fr/ffgenerator/.
\bibitem{ffs-recent} D. de Florian, R. Sassot, and M. Stratmann,
                          Phys. Rev. D {\bf 75}, 114010 (2007); 
                                       {\bf 76}, 074033 (2007);
                     T. Kneesch, B. A. Kniehl, G. Kramer, and I. Schienbein,  
                          Nucl. Phys. B {\bf 799}, 34 (2008);
                     S. Albino, B. A. Kniehl, and G. Kramer, 
                          Nucl. Phys. B {\bf 803}, 42 (2008); 
                     E. Christova and E. Leader, 
                          Phys. Rev. D {\bf 79}, 014019 (2009); 
                     S. Albino and E. Christova, 
                          Phys. Rev. D {\bf 81}, 094031 (2010). 
                     See also B. A. Kniehl and G. Kramer, 
                          Phys. Rev. D {\bf 74}, 037502 (2006).
\bibitem{Epele:2012vg} M.~Epele, R.~Llubaroff, R.~Sassot and M.~Stratmann,
                       Phys.\ Rev.\ D {\bf 86}, 074028 (2012);
    D. de Florian, R. Sassot, M. Epele, R. J. Hernandez-Pinto, and M.~Stratmann,
       Phys.\ Rev.\ D {\bf 91}, 014035 (2015);
    D. P. Anderle, M. Stratmann, and F. Ringer,
       Phys.\ Rev.\ D {\bf 92}, 114017 (2015).
\bibitem{ff-model} For example, see
                    B. Andersson, G. Gustafson, G. Ingelman, and T. Sjostrand,
                          Phys. Rept. {\bf 97}, 31 (1983);
                    Y. Hatta and T. Matsuo, Phys. Lett. B {\bf 670}, 150 (2008);
                    T. Ito, W. Bentz, I. C. Cloet, A. W. Thomas, and K. Yazaki,
                          Phys. Rev. D {\bf 80}, 074008 (2009);
        H. H. Matevosyan, A. W. Thomas, and W. Bentz,
              Phys. Rev. D {\bf 83}, 074003  (2011);
              D {\bf 83}, 114010 (2011), Erratum-ibid. D {\bf 86}, 059904 (2012);
        S. Nam and C.-W. Kao, Phys. Rev. D {\bf 85}, 034023 \& 094023 (2012).
        D.-J. Yang, F.-J. Jiang, C.-W. Kao, and S. Nam,
              Phys. Rev. D {\bf 87}, 094007  (2013).
        D.-J. Yang, F.-J. Jiang, W.-C. Chang, C.-W. Kao, and S. Nam,
              Phys. Lett. B {\bf 755} 393 (2016).
\bibitem{nuclear-ffs}
      X. F. Guo and X. N. Wang, Phys. Rev. Lett. {\bf 85}, 3591 (2000); 
      A. Majumder, E. Wang, and X. N. Wang, Phys. Rev. C {\bf 73}, 044901 (2006); 
      N. Armesto, L. Cunqueiro, C. A. Salgado, and W. C. Xiang, 
                    JHEP {\bf 02}, 048 (2008); 
      F. Arleo, Eur. Phys. J. C {\bf 61}, 603 (2009);
      S. Albino, B. A. Kniehl, and R. Perez-Ramos,
                    Nucl. Phys. B {\bf 819}, 306 (2009);
      A. Accardi, F. Arleo, W. K. Brooks, D. d'Enterria, and V. Muccifora, 
                    Riv. Nuovo Cimento, {\bf 32}, 439 (2010); 
      R. Sassot, M. Stratmann, P. Zurita, Phys. Rev. D {\bf 81}, 054001 (2010);
      W.-T. Deng and X.-N. Wang, Phys. Rev. C {\bf 81}, 024902 (2010);
      R. Dupr\'e, Ph. D. thesis, University of Lyon (2011).
\bibitem{hermes-nuclear} 
A. Airapetian {\it et al.} (HERMES collaboration), 
                           Nucl. Phys. B {\bf 780}, 1 (2007);
                           Phys. Lett. B {\bf 684}, 114 (2010);
                            	arXiv:1212.5407 [hep-ex].
\bibitem{belle} M. Leitgab {\it et al.} (BELLE Collaboration), 
                  Phys. Rev. Lett. {\bf 111}, 062002 (2013);
                M. Leitgab, Ph. D thesis,   
                University of Illinois at Urbana-Champaign (2013).
\bibitem{babar} J. P. Lees {\it et al.} (BABAR Collaboration), 
                  Phys. Rev. D {\bf 88}, 032011 (2013).
\bibitem{esw-book} R. K. Ellis, W. J. Stirling, and B. R. Webber,
             {\it QCD and Collider Physics}, Cambridge University Press (1996).
\bibitem{qqbar-cross} G. Altarelli, R. K. Ellis, G. Martinelli, and S. Y. Pi, 
                         Nucl. Phys. B {\bf 160}, 301 (1979);
              P. Nason and B. R. Webber, Nucl. Phys. B {\bf 421}, 473 (1994);
                         Erratum, {\it ibid.} {\bf 480}, 755 (1996).
              NNLO results are in P. J. Rijken and W. L. van Neerven,
                         Nucl. Phys. B {\bf 487}, 233 (1997);               
                       A. Mitov and S. Moch, 
                         Nucl. Phys. B {\bf 751}, 18 (2006).
\bibitem{ffs-def} J. Collins, Nucl. Phys. B {\bf 396}, 161 (1993);
                  G. Sterman {\it et al.}, Rev. Mod. Phys. {\bf 67}, 157 (1995).
\bibitem{splitting} 
  G.~Curci, W.~Furmanski, and R.~Petronzio,
          Nucl. Phys. B {\bf 175}, 27 (1980);
  W.~Furmanski and R.~Petronzio,
          Phys. Lett. B {\bf 97}, 437 (1980);
  E.~G.~Floratos, C.~Kounnas and R.~Lacaze,
          Nucl. Phys. B {\bf 192}, 417 (1981). 
   NNLO results are in  A. Mitov, S. Moch, and A. Vogt, 
                         Phys. Lett. B {\bf 638}, 61 (2006);
   S. Moch and A. Vogt, Phys. Lett. B {\bf 659}, 290 (2008);
   A. A. Almasy, S. Moch, and A. Vogt, Nucl. Phys. B {\bf 854}, 133 (2012).
\bibitem{space-time}
Relations between the spacelike and timelike splitting functions are
discussed in 
  M. Stratmann and W. Vogelsang,
                         Nucl. Phys. {\bf B496} (1997) 41;
  J.~Blumlein, V.~Ravindran, and W.~L.~van Neerven,
      Nucl. Phys. B {\bf 586}, 349 (2000);
  Y.~.L.~Dokshitzer, G.~Marchesini and G.~P.~Salam,
      Phys. Lett. B {\bf 634}, 504 (2006).
  See also articles in Ref. \cite{splitting}.
\bibitem{evolution} M. Miyama and S. Kumano,
                         Comput. Phys. Commun. {\bf 94}, 185 (1996);
                    M. Hirai, S. Kumano, and M. Miyama,
                         Comput. Phys. Commun. {\bf 108}, 38 (1998);
                                               {\bf 111}, 150 (1998);
                    S. Kumano and T.-H. Nagai, 
                         J. Comput. Phys. {\bf 201}, 651 (2004).
\bibitem{hk-q2evol} M. Hirai and S. Kumano,
                        Comput. Phys. Commun. {\bf 183}, 1002 (2012).
\bibitem{flavor3} S. Kumano, Phys. Rept. {\bf 303}, 183 (1998);
                  G. T. Garvey and J.-C. Peng,
                       Prog. Part. Nucl. Phys. {\bf 47}, 203 (2001);
	   J.-C. Peng and J.-W. Qiu, Prog. Part. Nucl. Phys. {\bf 76}, 43 (2014).
\bibitem{tasso12_30} R. Brandelik {\it et al.} (TASSO collaboration), 
                          Phys. Lett. B {\bf 94}, 444 (1980).
\bibitem{tasso14_22} M. Althoff {\it et al.} (TASSO collaboration), 
                          Z. Phys. C {\bf 17}, 5 (1983).
\bibitem{tasso34_44} W. Braunschweig {\it et al.} (TASSO collaboration), 
                          Z. Phys. C {\bf 42}, 189 (1989).
\bibitem{tpc29}      H. Aihara {\it et al.} (TPC collaboration), 
                          Phys. Rev. Lett. {\bf 52}, 577 (1984); 
                                           {\bf 61}, 1263 (1988).
\bibitem{hrs29}   M. Derrick {\it et al.} (HRS collaboration), 
                          Phys. Rev. D {\bf 35}, 2639 (1987).
\bibitem{topaz58} R. Itoh {\it et al.} (TOPAZ collaboration), 
                          Phys. Lett. B {\bf 345}, 335 (1995).
\bibitem{sld91}   K. Abe {\it et al.} (SLD collaboration), 
                          Phys. Rev. D {\bf 69}, 072003 (2004).
\bibitem{aleph91} D. Buskulic {\it et al.} (ALEPH collaboration), 
                          Z. Phys. C {\bf 66}, 355 (1995); 
                  R. Barate {\it et al.}, Phys. Rep. {\bf 294}, 1 (1998).
\bibitem{opal91}  R. Akers {\it et al.} (OPAL collaboration), 
                          Z. Phys. C {\bf 63}, 181 (1994).
\bibitem{delphi91} P. Abreu {\it et al.} (DELPHI collaboration), 
                          Eur. Phys. J. C {\bf 5}, 585 (1998).
\bibitem{delphi91-2} P. Abreu {\it et al.} (DELPHI collaboration), 
                          Nucl. Phys. B {\bf 444}, 3 (1995).
\bibitem{durham} http://hepdata.cedar.ac.uk/review/ee/.
\bibitem{resum} S. Albino, B.A. Kniehl, G. Kramer, and W. Ochs,
                          Phys. Rev. D {\bf 73}, 054020 (2006).
\bibitem{Furmanski:1981cw}  W.~Furmanski and R.~Petronzio,
                  Z. Phys. C {\bf 11}, 293 (1982).
\bibitem{hm-book} F. Halzen and A. D. Martin,
             {\it Quarks and Leptons: An Introductory Course 
                  in Modern Particle Physics}, John Wiley \& Sons (1984).
\bibitem{Beringer:1900zz} 
        J.~Beringer {\it et al.}  [Particle Data Group Collaboration],
                 Phys. Rev. D {\bf 86}, 010001 (2012).
\bibitem{th-mass} A. D. Martin, R. G. Roberts, W. J. Stirling, 
                  and R. S. Thorne,
                    Eur. Phys. J. C {\bf 23}, 73 (2002);
                    Phys. Lett. B {\bf 531}, 216 (2002).
\bibitem{chi2-ref} For example, see 
            G. D'Agostini, Nucl. Instrum. Meth. A {\bf 346}, 306 (1994);
            R. D. Ball {\it et al.}
               (NNPDF Collaboration), JHEP {\bf 05}, 075 (2010).
\bibitem{minuit} F. James, CERN Program Library Long Writeup D506 in
         https://root.cern.ch/sites/d35c7d8c.web.cern.ch \\ /files/minuit.pdf;
         see also https:www.cern.ch/minuit.
\bibitem{pdf-errors} J. Pumplin {\it et al.},
                         Phys. Rev. D {\bf 65} (2001), 014012 \& 014013;
               A. D. Martin {\it et al.},
                  Eur. Phys. J. C {\bf 28} (2003), 455; {\bf 35} (2004), 325.
\bibitem{errors} M. Hirai, S. Kumano, and N. Saito,
                  Phys. Rev. D {\bf 69} (2004), 054021;
                             D {\bf 74} (2006), 014015;
                 M. Hirai and S. Kumano, Nucl. Phys. B {\bf 813} (2009), 106;
                 M. Hirai, S. Kumano, and T.-H. Nagai,
                     Phys. Rev. C {\bf 70} (2004), 044905;
                                  {\bf 76} (2007), 065207. 
\bibitem{del-chi2} For example, see 
     https://cern-tex.web.cern.ch/cern-tex/minuit/node33.html,
     http://pdg.lbl.gov/2015 \\ /reviews/rpp2015-rev-statistics.pdf.
\bibitem{n-strange}   
          A. Airapetian {\it et al.} (HERMES Collaboration), 
                         Phys. Lett. B {\bf 666}, 446 (2008);
                         Phys. Rev. D {\bf 89}, 097101 (2014);
          W.-C. Chang and J.-C. Peng, 
                         Phys. Rev. Lett. {\bf 106}, 252002 (2011);
          H. Kawamura, S. Kumano, and Y. Kurihara, 
                         Phys. Rev. D {\bf 84}, 114003 (2011).  
\bibitem{sdis-delta-s} E. Leader, A. V. Sidorov, and D. B. Stamenov,
                         Phys. Rev. D {\bf 84}, 014002 (2011).
\bibitem{hkos08} M. Hirai, S. Kumano, M. Oka, and K. Sudoh,
                      Phys. Rev. D {\bf 77} (2008), 017504. 
\end{thebibliography}
\end{document}